\documentclass[11pt,a4paper,english]{article}
\usepackage{times,helvet}
\usepackage{ulem,vmargin}
\usepackage{amssymb}
\usepackage{color}
\usepackage{graphicx}
\usepackage[T1]{fontenc}
\input cyracc.def

\setpapersize{A4}

\author{Fran\c{c}ois Vernotte, Antoine Baudiquez, Enrico Rubiola}
\title{The spectrum decorrelation assumption\\ for the cross-spectrum method}
\date{October, 9,2020 (first version July 17, 2020)}
\begin{document}
\maketitle

\section*{Abstract}
This paper presents a very simple method ensuring the independence of consecutive spectra of the phase or frequency noise of an oscillator. This condition is essential for using cross-spectrum averages. 

\section{Introduction}
The best improvement provided by the cross-spectrum method \cite{rubiola10} arises from the average of consecutive cross-spectra. Such average relies on the decorrelation of the cross-spectra. Are the cross-spectra really uncorrelated? More generally, are not several consecutive spectra of the same process correlated?

\section{Theoretical approach}

\begin{figure}[!b]
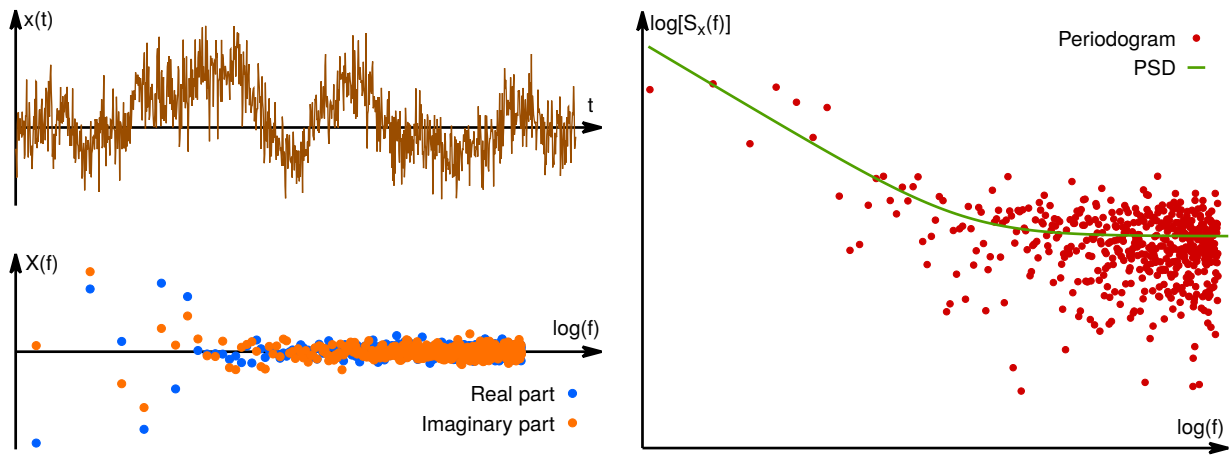

\begin{minipage}{0.48\linewidth}
\includegraphics[height=3.2cm]{time_data_simple_cinemascope.pdf}
\includegraphics[height=3.2cm]{spectrum_simple_cinemascope.pdf}
\end{minipage} \begin{minipage}{0.48\linewidth}
\includegraphics[height=6.4cm]{periodo_PSD_simple_4-3.pdf}
\end{minipage}
\caption{Time sequence (top left), spectrum (bottom left), periodogram and PSD (right).\label{fig:tf_PSD}}
\end{figure}

	\subsection{Reminder: The Fourier transform of a white noise is a white noise}
First of all, let us remind that the Fourier transform of a white noise is a white noise \cite{fessler98}. To be convinced, we may recall that the Fourier transformation is an orthogonal transformation which preserves the non-correlation of the data. Therefore, the periodogram of a white noise, i.e. the square of the modulus of its Fourier transform divided by the number of samples, is a sequence of $\chi^2$ random variables (r.v.) with 2 degrees of freedom: they are always positive and fluctuate around the white noise level but they remain fundamentally random. Only the Power Spectral Density (PSD) of a white noise is supposed to be a constant deterministic function since it is defined as a mathematical expectation. For this very reason, the PSD is an ideal mathematical quantity which can be only estimated (e.g. by an average of periodograms).

	\subsection{Filtering a white noise preserves the non correlation of the Fourier components\label{sec:filtering}}
A colored noise is generally considered as a filtered white noise. For instance, filtering a white noise with a transfer function $\propto 1/f$ yields a random walk, i.e. a noise characterized by a $f^{-2}$ PSD. Since filtering is nothing but a product by a deterministic function in the Fourier domain, such a process modulate the initial white spectrum but does not introduce any correlation between the frequency data. For example, the red noise of Fig. \ref{fig:tf_PSD}-top-left may be obtained by filtering a white noise. Thus, we retrieve in its Fourier transform (see Fig. \ref{fig:tf_PSD}-bottom-left) the uncorrelated complex r.v. of the white noise Fourier transform modulated by an envelope which is the transfer function of the filter. 

Therefore, the amplitudes of the spectrum of a pure colored noise are uncorrelated. Likewise, by extension, the spectra of 2 consecutive sequences of a pure colored noise are also uncorrelated. 

	\subsection{The residual mean of a sequence induces a drift after integration\label{sec:mean2drift}}
Let us consider a finite time series composed of $N$ uncorrelated centered Gaussian r.v. $z_k$ ($k \in \left\{1,N\right\}$), i.e. a white noise, with a standard deviation $\sigma$. Each r.v. follows then a Laplace-Gauss distribution that we will denote $\textrm{LG}(0,\sigma)$. Although the mathematical expectation of each r.v. is null, the arithmetic mean of the time series is not: it is a centered r.v. with a standard deviation $\sigma_0=\sigma/\sqrt{N}$
$$
z_0=\frac{1}{N}\sum_{k=1}^N z_k=\textrm{LG}(0,\sigma_0) \qquad \textrm{with} \qquad \sigma_0=\frac{\sigma}{\sqrt{N}}.
$$

If this white noise occurs in an integrator process of an oscillator, this constant $z_0$ will induce a linear drift of slope $z_0$. Considering a sampling time $\tau_0$, the following ``fake drift''\footnote{It is called  a ``fake drift'' in the sense that it acts as a true deterministic drift despite its random origin.} will be added to the integrated time series: $d_k=z_0 k\tau_0$.

Admittedly the mathematical expectation of the slope $z_0$ decreases as $1/\sqrt{N}$, but its influence over the whole sequence increases as $N$: at the end of the time series, the mathematical expectation of the drift reaches 
$$
\mathbb{E}[d_N]=\frac{\sigma}{\sqrt{N}}N\tau_0=\sigma\sqrt{N}\tau_0.
$$

Therefore, the influence of the fake drift increases as the square root of the number of elements in the time series. For large $N$, this can even overwhelm the random behavior of the integrated time series.

	\subsection{Only deterministic behaviors induce correlations}
Now, let us come back to the main issue of this paper: are the cross-spectra of consecutive time series uncorrelated? une méthode assurant l'indépendanceFrom {\S} \ref{sec:filtering}, we can answer that the random part of these cross-spectra are uncorrelated. On the other hand, if the filtering corresponds to an integration (transfer function $\propto 1/f$ or $1/f^n$ for $n^\textrm{\footnotesize th}$ order integration), a fake drift will appear as described in {\S} \ref{sec:mean2drift}. This drift will add a deterministic component in the spectrum (see \cite{vernotte19} for a full description of this phenomenon) which will induce strong correlations between the frequency data as well as between the spectra of consecutive time series\footnote{Throughout this paper,  ``consecutive time series'' means consecutive subdivisions of the whole time series.}.

Therefore, such an effect completely prohibits the use of consecutive cross-spectra if a drift, fake or real, spans several consecutive time series.

	\subsection{Syntonizing: a simple but effective way to decorrelate spectra\label{sec:synton}}
Nevertheless, the remedy for this major failure is quite obvious: we just have to break the correlations between spectra of consecutive time series by removing the linear drift of each consecutive time series or, and this is perfectly equivalent, by forcing the mean of each consecutive time series to 0 before integration.

But, since we are only interested in the decorrelation of the spectra of consecutive time series and not in the decorrelation of the frequency data within a spectrum, a simpler way exists: it suffices to force the first data of each time series to 0 before integration. In this case, there will still be a residual drift in each time series after integration but all these residual drifts will be uncorrelated from one spectrum to another.
If we consider that the final time series corresponds to phase-time data, this means that the initial time series before integration are frequency deviation data. This forcing at 0 corresponds therefore to a syntonization process at the beginning of each consecutive time series.

\section{Monte-Carlo approach}
We will first simulate the case of a Random Walk FM noise ($f^{-2}$ frequency noise $\equiv$ $f^{-4}$ phase noise) and then extend to other types of noises.
 
	\subsection{Red noise simulation\label{sec:simul}}
\begin{figure}
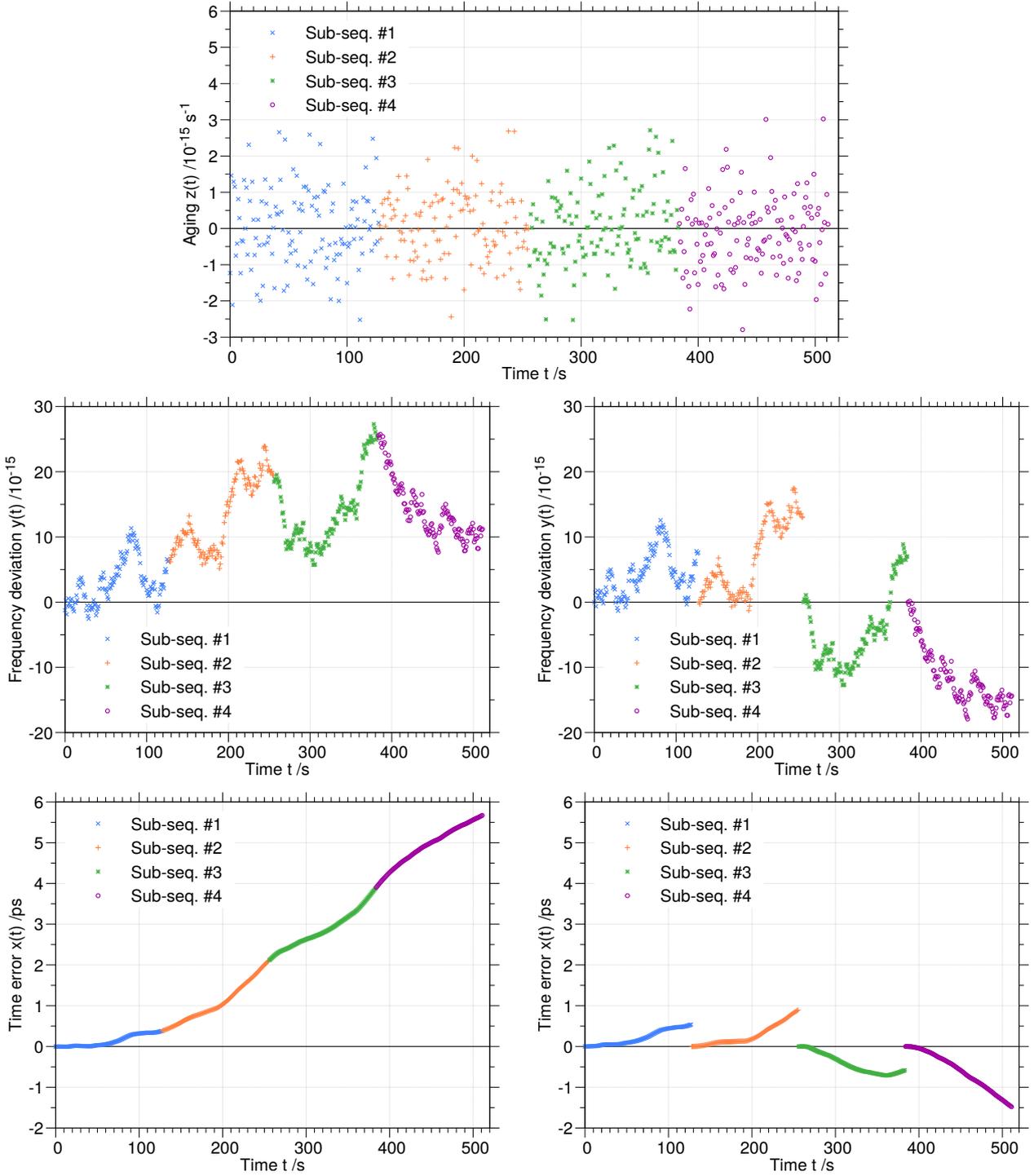

\centering
\includegraphics[height=6.4cm]{Z_sub_seq_ref.pdf}

\raggedright
\includegraphics[height=6.4cm]{Y_no_synt_ref.pdf}\includegraphics[height=6.4cm]{Y_synton_ref.pdf}

\includegraphics[height=6.4cm]{X_no_synt_ref.pdf}\includegraphics[height=6.4cm]{X_synton_ref.pdf}

\caption{From $z(t)$ to $x(t)$ without (left) or with (right) syntonisation.\label{fig:z2x}}
\end{figure}

In order to simulate a $f^{-4}$ red noise, we will integrate twice a white noise. Let us define the aging $z(t)$ as a Laplace-Gauss r.v. of mathematical expectation 0 and of standard deviation 1: $z(t_k)=\textrm{LG}(0,1)$ with $k\in\left\{1, \ldots, N\right\}$ and $t_{k+1}-t_k=\tau_0$, the sampling time. For the sake of simplification, we set $\tau_0=1$ s and we denote $z_k=z(t_k)$. The top graph of Figure \ref{fig:z2x} represents a sequence of 512 consecutive realizations of such a r.v..

Although $z(t)$ is discontinuous at the scale given by $\tau_0$, let us assume that $z(t)$ is smooth and continuous for very short term, i.e. at a scale $\epsilon\ll \tau_0$ (e.g. $z(t)$ could be a first order Markov process with a time constant $\epsilon$ \cite{kasdin1995}). Thanks to this assumption, this process has a finite power and its Fourier transform $Z(f)$ exists. 

Since aging is defined as the time derivative of the frequency deviation $y(t)$, $z_k=\frac{y(t_k+\tau_0)-y(t_k)}{\tau_0}$, we can compute the mean frequency deviation between $t_{k-1}$ and $t_k$, that we denote $\bar{y}_k=\frac{1}{\tau_0}\int_{t_{k-1}}^{t_k}y(t)\textrm{d}t$, as the cumulative sum of the aging samples $z_k$ from the beginning of the sequence:
$$
\bar{y}_k=\bar{y}_{k-1}+z_k \tau_0=\tau_0\sum_{j=1}^k z_j.
$$
The left hand graph in the middle of Figure \ref{fig:z2x} represents the corresponding 512 frequency deviation samples.

Similarly, we can deduce the phase-time data by integrating the frequency deviation data-run:
$$
x(t_k)=\int_0^{t_k}y(t)dt=\tau_0\sum_{j=1}^k \bar{y}_j.
$$
The result of this process is plotted on the left hand graph at the bottom of Figure \ref{fig:z2x}.

As stated in {\S} \ref{sec:mean2drift}, the slight residual mean of the aging sequence induces a linear drift in the frequency deviation sequence which induces a quadratic drift in the phase-time sequence. The 512 data sequence are divided in 4 subsequences of 128 data (see blue, red, green and purple dots in Fig. \ref{fig:z2x}). It is clear that this effect, and then the correlations between subsequences, increases drastically with the order of the subsequence.

On the other hand, as expected in \ref{sec:synton}, the syntonization breaks the correlations between the subsequences. The right hand side of Fig. \ref{fig:z2x} shows the syntonized frequency deviation sequence (middle) and the induced phase-time sequence (bottom). The frequency deviation subsequences are the same as the ones without syntonization except that they are shifted in such a way that the first sample is null. The changes in the phase-time subsequences are a little bit more significant since, in addition to the shift of the subsequences, there is also a change of slope due to a change of the mean-value of the frequency deviation subsequences.

	\subsection{Spectrum correlation analysis}
\begin{figure}
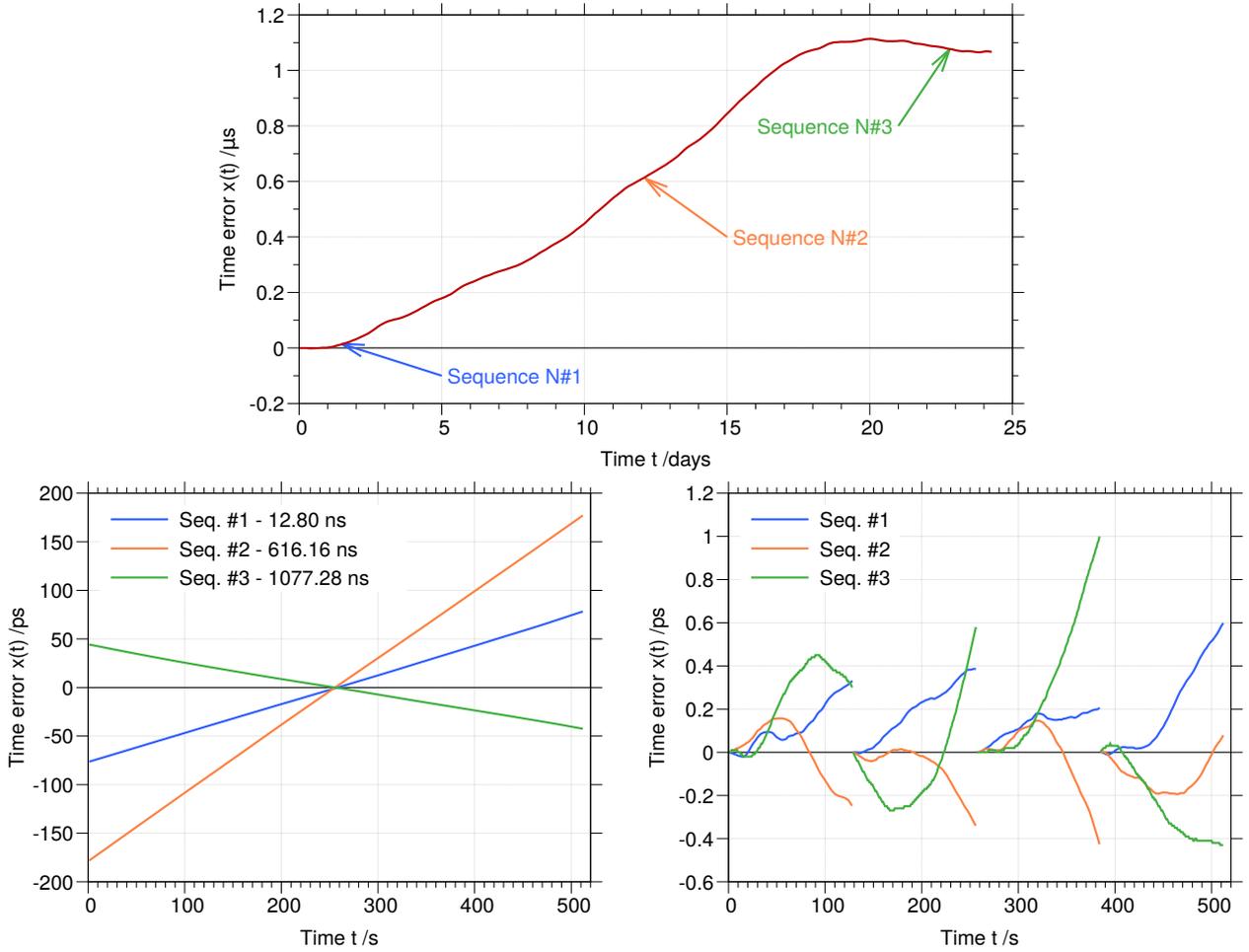

\centering
\includegraphics[height=6.4cm]{X_rough_ref.pdf}

\raggedright
\includegraphics[height=6.4cm]{X_3_subseq_slopes_ref.pdf}\includegraphics[height=6.4cm]{X_3_subseq_synton_ref.pdf}

\caption{Definition of 3 sequences and 4 subsequences without (left) or with (right) syntonisation.\label{fig:subseq}}
\end{figure}

In order to analyze the correlation between the subsequences spectra, we applied the method described in {\S} \ref{sec:simul} to realize a $f^{-4}$ red noise data-run with $2\,095\,152$ samples, i.e. $16384$ subsequences $\times 128$ data. Fig. \ref{fig:subseq}-top represents such a realization with a sampling period of $1$ s and then a whole length of $T \approx 24$ days.

Within this whole continuous data-run, we define $3$ sequences each containing 4 sub-sequences of 128 data:
\begin{itemize}
	\item Sequence \#1 is close to the beginning (from samples $128\,000$ to $128\,511$)
	\item Sequence \#2 is close to the center (from samples $1\,048\,576$ to $1\,049\,087$)
	\item Sequence \#3 is close to the end (from samples $1\,969,152$ to $1\,969\,663$).
\end{itemize} 

Fig. \ref{fig:subseq}-middle-left shows these 3 sequences in the same plot, i.e. an enlargement of the 3 areas pointed by the arrows in Fig. \ref{fig:subseq}-top. The main features are drifts which range over a few tens of ps over each sequence.

On the other hand, if the syntonization process is done at the beginning of each 128-data subsequence, the curve appearance is quite different (see \ref{fig:subseq}-middle-right) and is limited to less than $1$ ps peak-to-peak: the linear drift is much lower revealing other features like random behavior as well as quadratic drifts.

\begin{figure}
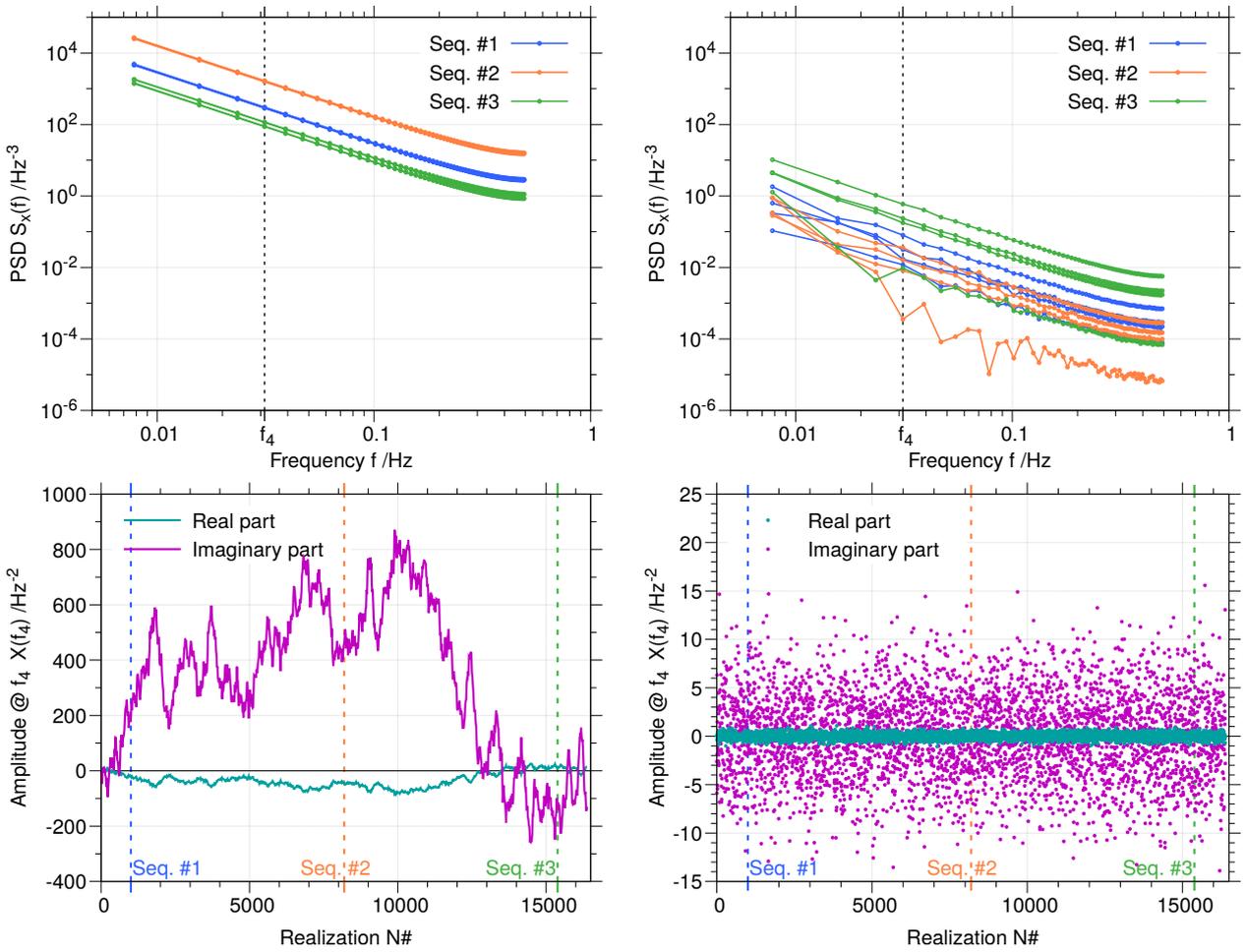

\raggedright
\includegraphics[height=6.4cm]{3X5_PSD_no_synt_ref_alt.pdf}\includegraphics[height=6.4cm]{3X5_PSD_synton_ref_alt.pdf}

\includegraphics[height=6.4cm]{ampl_4th_f_no_synt_ref.pdf}\includegraphics[height=6.4cm]{ampl_4th_f_synton_ref.pdf}
\caption{PSD (above) and realizations of the $4^\textrm{\footnotesize th}$ amplitude of the spectrum (below) without (left) or with (right) syntonisation.\label{fig:spec}}
\end{figure}

We computed the periodograms of the $16\,384$ subsequences of the rough and syntonized data-runs. Fig. \ref{fig:spec}-top shows the periodograms of $3\times 4$ subsequences defined above. Without syntonization (left), the periodograms are smooth and do not follow the expected $f^{-4}$ slopes. Although their shapes are very similar, their levels are quite different depending of the sequence N\#, e.g. the level of Sequence \# 2 is about $15$ times higher than the one of Sequence \# 3. These features have to be related to the time curves of Fig \ref{fig:subseq}-middle-left: we observe the periodograms of strong linear drifts which completely mask the random behavior of these subsequences.

With syntonization (see Fig. \ref{fig:spec}-top-right), The periodograms are much lower and the random behavior becomes visible. It is particularly clear on the lowest orange curve which exhibits a $f^{-4}$ trend from $0.01$ to $0.1$ Hz.

The lower part of Fig. \ref{fig:spec} focuses on the behavior of one of the amplitudes of the Fourier transform of all $16\,384$ subsequences. We arbitrarily chose the amplitude at $f_4=\frac{4}{128\tau_0}=31.25$ mHz. In other words, the sum of the square of the real part plus the square of the imaginary part divided by $128$ at the blue dotted line gives the amplitude of the blue periodogram at $f_4$.  Fig. \ref{fig:spec}-bottom-left shows clearly that the real parts are strongly correlated from one subsequence to another as well as the imaginary parts. On the other hand, Fig. \ref{fig:spec}-bottom-right shows that with syntonization these correlations seems to disappear. But before to conclude this theoretical study, let's analyze more thoroughly the correlations of these data.

\begin{figure}
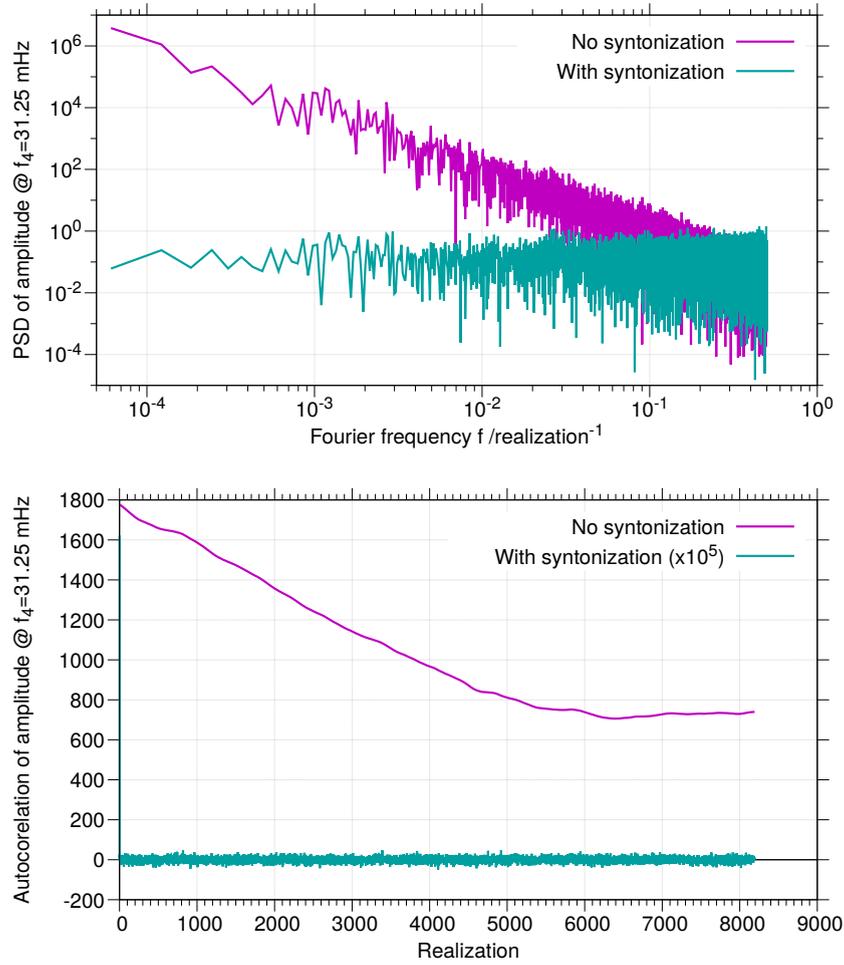

\centering
\includegraphics[height=6.4cm]{PSD_4th_f_ref_alt.pdf}

\includegraphics[height=6.4cm]{Autocor_4th_f_ref_alt.pdf}
\caption{PSD (above) and autocorrelation (below) of the $4^\textrm{\footnotesize th}$ amplitude of the spectrum without (purple) or with (cyan) syntonisation.\label{fig:correl}}
\end{figure}

In order to characterize these correlations, let us remind that, for each subsequence, its Fourier transform at $f_4$ is a complex r.v. The issue consists then to observe how these r.v. are correlated versus the time shift of the corresponding subsequences. E. g., with syntonization, \ref{fig:spec}-bottom-left shows a strong correlation between neighboring sequences. Therefore, we ought to estimate the autocorrelation of all these complex r.v. versus the N\# of the realization (i.e. the N\# of the sequence). The easier way to perform such an estimation consists in computing the periodogram of these complex r.v. (the periodogram of periodogram amplitudes at $f_4$!) and then compute its inverse Fourier transform. This is what is presented on Fig. \ref{fig:correl}: above the periodograms of the amplitudes at $f_4$ without (purple) or with syntonization (cyan), below the corresponding autocorrelation functions. These results confirm what we guessed at the previous paragraph: with syntonization, the autocorrelation function tends toward a Dirac peak and then the complex r.v. are uncorrelated. On the other hand, it is interesting to notice that, without syntonization, this complex r.v. follows a random walk noise. 

By using a similar study on $f^-3$ red noise, we remarked that the amplitude at $f_4$ without syntonization is also a complex r.v. but follows a flicker noise. For the other 3 types of noise generally encountered in time and frequency metrology (white FM, flicker PM and white PM), syntonization is innocuous though useless.

\section{Experimental approach}
Experimental measurements are in progress to validate these theoretical results.

\section{Conclusion}
This study proves that a simple syntonization of an oscillator breaks the memory of the non stationary processes which are in its phase or frequency noise and then, gives uncorrelated spectra. Using this property allows us to average all the cross-spectra obtained after syntonization in order to converge to the phase or frequency PSD of the oscillator.

\bibliography{crospec}
\end{document}